\documentclass{nature}
\usepackage{epsfig}
\usepackage{amsmath}
\usepackage{amssymb}

\begin{document}

\title{The d-p band-inversion topological insulator in bismuth-based skutterudites}
\author{Ming Yang and Wu-Ming Liu$^{\star}$}
\maketitle

\begin{affiliations}
\item
Beijing National Laboratory for Condensed Matter Physics,
Institute of Physics, Chinese Academy of Sciences,
Beijing 100190, China

$^\star$e-mail: wliu@iphy.ac.cn

\end{affiliations}

\begin{abstract}
%Topological insulator, a class of material with protected gapless surface states and a gapped bulk energy gap, is proposed to host fascinating phenomena such as Dirac Fermions, Majorana Fermions, quantum anomalous Hall effect as well as topological magnetoelectric effect, and hence receives considerable research interests.
%Topological phase transition connects the topological trivial states with those exotic states such as the topological insulating state, the quantum anomalous Hall state, topological crystalline insulating state etc., and hence received warm research interest.
Skutterudites, a class of materials with cage-like crystal structure which have received considerable research interest in recent years, are the breeding ground of several unusual phenomena such as heavy fermion superconductivity, exciton-mediated superconducting state and Weyl fermions.
%Strain engineering provides a promising way to control the band structures of materials and even their topological property.
Here, we predict a new topological insulator in bismuth-based skutterudites, in which the bands involved in the topological band-inversion process are d- and p-orbitals, which is distinctive with usual topological insulators, for instance in Bi$_2$Se$_3$ and BiTeI the bands involved in the topological band-inversion process are only p-orbitals. Due to the present of large d-electronic states, the electronic interaction in this topological insulator is much stronger than that in other conventional topological insulators.
The stability of the new material is verified by binding energy calculation, phonon modes analysis, and the finite temperature molecular dynamics simulations.
This new material can provide nearly zero-resistivity signal current for devices and is expected to be applied in spintronics devices.
\end{abstract}

%\section*{Introduction}

Topological insulator (TI) is a new kind of material which has gapped bulk state and gapless surface state with the latter protected by the topological character of TI\cite{BHZ,Heusler_SCZhang,magnetoelectric,Yan1,Fang_Bi2Se3,Fu,YXia,silicene_Yao,SunFD}. The topological order parameter of TI in general cases is spin Chern number, and TI under time reversal symmetry is characterized by Z$_2$ quantum number\cite{Fu}. The unique features of its surface state make TI have potential applications in spintronics and quantum information devices. TI is also the breeding ground for a good number of interesting quantum phenomena such as quantum anomalous Hall effect\cite{QAHE,ZHQiao,Ding,ZhangXL,JianminZhang}, Majorana fermions\cite{Majorana_Fu,Majorana_Tiwari} and topological magnetoelectric effect\cite{magnetoelectric}.
TIs usually appear in those materials containing elements with strong spin-orbit coupling, for example, the bismuth element in Bi$_2$Se$_3$\cite{Fang_Bi2Se3,weizhang}, BiTeI\cite{Bahramy,GuoGuangYu}, and ScPtBi\cite{Heusler_SCZhang}.
%Despite TIs in natural state, people also find the so-called "Pressure-induced" Topological Insulators.
Moreover, pressure and strain has been demonstrated as an effective way to modulate the topological property of materials. For instance, CdSnAs$_2$ under a 7\% decrease in the lattice constant will become topological insulator\cite{Yao_Chalcopyrite} while a 6\% change in the length of c-axis will drive Bi$_2$Se$_3$ from topological non-trivial phase into topological trivial phase\cite{xiangtan}. However, such TIs with p-band inversion are free of strong correlation effects and thus do not possess those interesting phenomena induced by electronic interaction, such as the transition in correlated Dirac fermions\cite{XieXC} and interaction induced topological Fermi liquids\cite{Castro}.
Consequently, those TIs beyond p-band inversion arouse intensive research interest.
Here, we predict a new d-p band-inversion topological insulator in bismuth-based skutterudites in which the bands involved in the topological band-inversion process are d- and p-orbitals. Due to the present of large d-electronic states, the electronic interaction in this topological insulator is much stronger than that in other conventional topological insulators.
%
%
%Topological insulator is a new kind of material which has gapped bulk state and gapless surface state with the latter protected by TI's topological character. The topological order parameter of TI in general cases is Spin Chern Number, and TI under time reversal symmetry is characterized by Z$_2$ quantum number\cite{Fu}. The unique features of its surface state make TI have potential applications in Spintronics and quantum information devices.
%TIs usually appear in the material containing elements with strong spin-orbit coupling, for example, the bismuth element in Bi$_2$Se$_3$\cite{Fang_Bi2Se3}, BiTeI\cite{Bahramy}, and ScPtBi\cite{Heusler_SCZhang}.
%%Despite TIs in natural state, people also find the so-called "Pressure-induced" Topological Insulators.
%Moreover, pressure and strain has been demonstrated as an effective way to modulate the topological property of materials. For instance, CdSnAs$_2$\cite{Yao_Chalcopyrite} under a 7\% decrease in the lattice constant will become topological insulator while a 6\% change in the length of c-axis will drive Bi$_2$Se$_3$ from topological non-trivial phase into topological trivial phase\cite{xiangtan}. However, in those TIs the bands involved in the topological band-inversion process are only p-orbitals. Here, we predict a new bismuth-based skutterudite in which the bands involved in the topological band-inversion process are d- and p-orbitals.

Skutterudites, such as RhAs$_3$, IrAs$_3$, IrSb$_3$ and IrP$_3$, crystallize in a cage-like crystal structure in which each transition metal atom octahedrally coordinates to six pnictide atoms\cite{RuSb3,Niwa} (see Fig. 1 (a)). They have large Seebeck coefficients and therefore can behave as excellent thermoelectric materials\cite{CoSb3}. The discovery of heavy fermion superconductivity\cite{PrOs4Sb12_Seyfarth}, exciton-mediated superconducting state\cite{PrOs4Sb12_Matsumoto} and Weyl fermions\cite{Pickett1} in this system makes skutterudites a hot spot in condensed matter physics. Besides those skutterudites naturally exist, a number of new members in skutterudites have been experimentally synthesized, such as  NiSb$_3$\cite{NiSb3} in 2002 and RuSb$_3$\cite{RuSb3} in 2004. However, those materials are composed of elements with relatively weak spin-orbit coupling (SOC). Knowing that topological insulators are usually those materials containing elements with strong spin-orbit coupling strength, such as the bismuth element in topological insulator Bi$_2$Se$_3$\cite{Fang_Bi2Se3,weizhang}, BiTeI\cite{Bahramy,GuoGuangYu} and LaPtBi\cite{Heusler_Yao}, people are reasonable to ask whether or not skutterudites composed of elements with strong spin-orbit coupling strength, i.e. bismuth, can exist stably and whether they can be topologically non-trivial? This new topological insulator in bismuth-based skutterudites, is exactly such kind of skutterudite materials which is able to exist stably, contains elements with strong SOC, and has controllable topological phase transition.

%Compression behaviors of Binary skutterudite£¬taking CoP$_3$ as an example, have been experimentally investigated in 2011. XRD results indicate that the structure remains stable up to as high as 40.4 GPa at room temperature\cite{Niwa} .

%In this letter, we firstly predict a new member in skutterudite family, bismuth-based skutterudite and
%For this material, we verified its stability
In this work, we predict a new d-p band inversion topological insulator in bismuth-based skutterudites, which is distinctive with usual topological insulators, for instance in Bi$_2$Se$_3$ and BiTeI the bands involved in the topological band-inversion process are only p-orbitals. Due to the present of large d-electronic states, the electronic interaction in this topological insulator is much stronger than that in other conventional topological insulators. The stability of the new material is verified by binding energy calculation, phonon modes analysis, and the finite temperature molecular dynamics (FTMD) simulations. We demonstrate that external strains are able to induce a topological phase transition in this system via band structure calculations. We confirm its topological non-trivial property by $Z_2$ quantum number calculation.
%Moreover, we give a topological phase diagram with parameters of SOC strength and on-site repulsion energy.

\section*{Results}

\subsection{Crystal structure and optimized lattice parameter.}
The bismuth-based skutterudite IrBi$_3$ investigated here has space group $IM\overline{3}$, and its crystal structure is shown in Fig. 1.
There are 8 Ir atoms and 24 Bi atoms in a unit cell. Each Ir atom is surrounded by 6 Bi atoms and each Bi atom has 2 Ir nearest neighbors (see Fig. 1 (a)). The structure has space inversion symmetry with the inversion center (1/2,1/2,1/2). The structure belongs to the body-centered lattice type, and its primitive cell (Fig. 1 (b)) has a half volume of the unit cell. Fig. 1 (c) shows the Brillouin zone and high symmetric points with $\Gamma$ (0,0,0), H (0,1/2,0), N (1/4,1/4,0), P (1/4,1/4,1/4).

We first optimize the lattice parameter and ionic positions. The calculated total free energy (solid line) as a function of lattice parameter is shown in Fig. 1 (d). It can be clearly seen that the optimized lattice parameter (corresponding to the position of free energy minimum) of the primitive cell is 8.493{\AA}. This value is 6 \% larger than that of IrSb$_3$\cite{IrSb3}, which can be explained that Bi atom has a larger atomic radius than Sb atom.

\subsection{Binding energy calculation, phonon modes analysis and the finite temperature molecular dynamics simulations.}
In order to verify the stability of the new material, the authors perform the binding energy calculation, phonon modes analysis and the finite temperature molecular dynamics (FTMD) simulations. The binding energy is calculated by
\begin{equation} E_{b}=E_{IrBi_3}-n_{Ir}{\cdot}E_{Ir}-n_{Bi}{\cdot}E_{Bi}, \end{equation}
where $E_{IrBi_3}$ denotes the free energy of IrBi$_3$ per primitive cell, $E_{Ir}$ and $E_{Bi}$ the free energy of crystalline Ir and Bi per atom. $n_{Ir}$ and $n_{Bi}$ the number of Ir and Bi atoms in IrBi$_3$ primitive cell. By simple calculation\cite{BindingEnergy},
$E_{b}$ is found to be equal to $-3.65$ eV per primitive cell. The negative value of binding energy infers a stable state of IrBi$_3$.

Fig. 2 shows the phonon dispersion and phonon density of states (DOS) for IrBi$_3$ at zero strain. In the phonon DOS subfigure, the black solid line represents the total phonon density of states, while the green and red shaded areas represent the states coming from Ir and Bi atoms, respectively. Phonon states in the low energy range are mostly composed of states from Bi atoms, indicating that Bi atoms in IrBi$_3$ are much easier to vibrate than the Ir atoms. The phonon dispersion and phonon DOS show no imaginary frequency, indicating that IrBi$_3$ is stable.

In addition, the dynamical stability of the material is further checked by finite temperature molecular dynamics simulations at temperature 300 K for room temperature and 30 K for low temperature. During the simulations, a $2\times2\times2$ supercell containing 256 atoms is used. The length of time-step is chosen as 5 fs and simulations with 1000 steps are executed.
It is observed that, the atoms shake around the equilibrium positions back and forth while the extent of such motion under 300 K is larger than under 30 K (the evolution of atomic positions can be found in movies in supplementary materials).
However, no structural collapse happens throughout the simulations, which can also be seen from the free energies curves as the functions of time-step shown in Fig. 3. It is also observed that, the crystal structure always remains nearly the same as the initial crystal structure. Actually, as is shown in the inset of Fig. 3, the crystal structure corresponding to the last free energy maximum in T=300 K case (right), still shows no significant structural differences as compared with the initial crystal structure (left).
%The inset of Fig.4 compares the initial crystal structure(left) with the crystal structure after 1000 steps in T=30K case(right), corroborating the inference discussed above.
%which provide another evidence that such material
%The free energies as the functions at each simulation step shown in Fig.4 shows  .
The lattice relaxation, binding energy calculation, phonon modes analysis together with FTMD simulations mentioned above provide an authentic test for the stability of bismuth-based skutterudite IrBi$_3$.

\subsection{Strain-induced d-p band-inversion topological insulator.}

The calculated band structures are listed in Fig. 4, where the black and blue lines represent the GGA and GGA+U band structures, respectively. As is shown in Fig. 4 (a), before exerting pressure, IrBi$_3$ resides in the normal metal state with its bands crossing the Fermi level several times. With the increase of the isotropic strain, its density of states (DOS) at Fermi level gradually decreases. Under a 9\% isotropic strain, the bands go across the Fermi level at $\Gamma$ point but not at other points in Brillouin Zone (BZ) (see Fig. 4 (b)), and the conduction band minimum and valence band maximum degenerate and the material behaves as a semi-metal which have a zero energy gap, just like Graphene and CeOs$_4$As$_{12}$\cite{Yan1}. This degeneracy at $\Gamma$ is protected by the cubic symmetry of crystal, which, as is tested by us, cannot be eliminated by small changes of the lattice constant. In order to shift the degeneracy at $\Gamma$, one needs to break that symmetry. An unsophisticated way is to add an anisotropy just like what was done on CdSnAs$_2$\cite{Yao_Chalcopyrite}. Here, we simply further impose a 2\% suppression on the c-axis of the primitive cell while remaining the length of a- and b-axis unchanged, which imposes anisotropy on the system. In this situation, it can be clearly seen in Fig. 4 (c) that there is a gap opened at the Fermi level, dragging the system in the insulating state.
%As the topological non-trivial properties in general case comes from the inversion of band order, we are eager to check whether the system are now has a normal band order or it have an inverted band order. To achieve this, we employ an orbital projection on p electron of Ir atom.
Fig. 4 (d) shows the Ir-d projected band structure near the Fermi level and near $\Gamma$ point, in which the radii of red circles correspond to the proportion of Ir-d electrons. It can be seen that,
%the highly dispersive band near the Fermi level is mainly contributed by p-orbitals of Bi atoms, while
those localized bands above the Fermi level are mainly contributed by d-orbitals of Ir atoms.
Such band-inversion character is further checked by the modified Becke-Johnson (mBJ) potential (see supplementary information), which is proved to be able to predict an accurate band gap and band order\cite{mBJ_JCP,mBJ_TranBlaha,mBJ_WXFeng}.
%This technique has been demonstrate an effective way to identify the band order\cite{Yan1}.
The highly dispersive band below the Fermi level is mainly contributed by p-orbitals of Bi atoms, and it has little weight of Ir atoms in those k-points far away from $\Gamma$ point. However, in the vicinity of $\Gamma$ point, the weight of Ir atoms in that band increases rapidly and becomes dominating orbital component, showing an apparent band inversion.
In order to further confirm the topological property in such condition, we calculate the $Z_2$ topological quantum number of the system by the Fu-Kane method\cite{Fu}. The index for strong topological insulators $v_{0}$ is expressed as $(-1)^{v_{0}}=\prod_{i=1}^{8}\delta_{i}$
 in which $\delta_{i}=\prod_{m=1}^{N}\xi_{2m}(\Gamma_{i})$ represents the product of the parities of each occupied band at 8 time-reversal invariant momenta $\Gamma_{i}$.
%Parity analysis has been shown to be an efficient way to determine topological property of systems with space inversion symmetry (SIS) \cite{Fang_Bi2Se3,weizhang,xiangtan}.
%Because there is still SIS in IrBi$_3$ even under bi-axis strain, the Z2 topological quantum number of the system can be calculated by parities of occupied bands at each time-reversal invariant momenta (TRIM), which was proposed by Fu and Kane\cite{Fu}.
The calculated parities of top-most isolated valence bands (here refers to the isolated block of states between -8.0 eV to 0 eV in Fig. 4) at eight time-reversal invariant momenta are listed in Table 1, where the deeper states (those states lower than -9.5 eV in Fig. 5) separated far from top-most isolated valence bands
%by a gap as large as approximately 2eV
are ignored because they don't change system's band topology. As is shown, the product of parities of occupied bands contributes a $-1$ at $\Gamma$ while $+1$ at the seven other time-reversal invariant momenta. As a result, $Z_2$ quantum number is $\nu_{0}=1$, $\nu_{1}=\nu_{2}=\nu_{3}=0$, which
corresponds to a strong topological insulator.

\subsection{Partial-density of states and the d-p orbitals dominating property near the Fermi level.}
Fig. 5 depicts the atomic- and orbital-resolved density of states (DOS). The black solid lines in subfigure (a)(b)(c) represent the total DOS. Fig. 4 (a) is atomic-resolved DOS, in which the green curve represents the states of Ir while the red curve represents the states of Bi. It's clear that the DOS of both types of atoms is in quite large values, indicating that both types of atoms make a significant contribution to the total DOS. This is different from MoS$_2$ where states near Fermi level are dominated by only one kind of atom (Mo)\cite{wanxiang_feng_MoS2}.
% different from those materials (such as MoS$_2$) where states near Fermi level are dominated by only one type of atom(Mo)\cite{wanxiang_feng_MoS2}.
Fig. 5 (b) and (c) are orbital-resolved DOS of Ir and Bi atoms, respectively. Green, blue and red curves represent s-, p- and d-orbitals. One character of the material introduced here is a large proportion of d-states near $E_{F}$.

Further, we calculate the d-orbitals projected band structures (see Fig. 6) in the local coordinate of the Bi octahedral. The orange, violet, red, green and blue colors in Fig. 6 represent the d$_{z^2}$, d$_{x^2-y^2}$, d$_{xy}$, d$_{yz}$ and d$_{xz}$ orbitals respectively. The radii of circles are proportional to the weights of corresponding orbitals. It can be seen that, the t$_{2g}$ orbitals (including the d$_{xy}$, d$_{yz}$ and d$_{xz}$ orbitals) reside far below the Fermi level and are fully occupied. While, the lowest three conduction bands are mainly contributed by the e$_g$ orbitals (including the d$_{z^2}$ and d$_{x^2-y^2}$ orbitals). More specifically, the d$_{x^2-y^2}$ orbital makes an even larger contribution than the the d$_{z^2}$ orbital for the lowest conduction band. The large proportion of d-states near $E_{F}$ is distinctively different from usual TI materials, for example states near $E_F$ mainly containing s-p electrons in HgTe and p-electrons in Bi$_2$Se$_3$.
The large proportion of d-states near $E_{F}$ indicates that electrons in such material process strong electronic correlations and are more localized than other TI materials.
The strong electronic correlations make the material a good platform for investigating the effect of correlations on the topology, as well as a candidate for realizing the quantum information device based on correlations.
The localization will enhance the effective mass of bulk electrons, and hence reduce the bulk contribution to the local current at finite temperature, making the spin-binding property more apparent, which is helpful in fabricating spintronics devices with higher stability.

%\subsection{Topological surface states.}
%%In order to further confirm the topology of the system, one need to calculate the surface band structure.
%Then, we calculated the surface band structure of the material.
%Based on the results of the first-principles calculation, we constructed a tight-binding model of a slab consisting 19 layers of primitive cell(arranged along c-axis) via Maximally Localized Wannier Function(MLWF) approach\cite{wannier1,wannier2} introduced by Marzari et al, and calculated surface band structure is shown in FIG.4.
%%From partial density of state shown in FIG.3 we can clearly see that the states near the Fermi level mainly come from the contribution of p orbitals of Bi and d orbitals of Ir. Consequently, projections are made on these dominant orbitals.
%%we  In the process of Wannier interpolation, projections were make on Ir-d and Bi-p orbitals, which were dominant ingredients near the Fermi level.
%%A slab consisting 19 layers of primitive cell(arranged along c axis) were used in such process. FIG.4 shows its surface states.
%In FIG.4, the $\Sigma_1$ band which starts from the valence band at $\Gamma$ point and inclines to conduction band, crossing the Fermi level 1(odd) time in half of BZ, is the topological surface band.
%%Other surface states($\Sigma_2$, $\Sigma_3$ and $\Sigma_4$) which cross Fermi level even times in half of BZ, don't change the topology of the system\cite{Kim}.

\section*{Discussion}
Experimentally, the new strain-induced topological insulator IrBi$_3$ could be grew using the Bridgman method, by which the CoP$_3$\cite{Niwa} and the RuSb$_3$\cite{RuSb3} crystals have been successfully synthesised. The crystal growth should be conducted in a sealed quartz ampoule. The iridium and bismuth should be coated by graphite and then introduced into the quartz ampoule. A temperature gradient of about 50$^{\circ}$C should be maintained at the growth interface, just like in the case of RhSb$_3$\cite{Caillat}. To remove the excess bismuth in the as-grown crystal, post-annealing should be performed\cite{annealing}.
After the synthesis of the new material, its crystal structure could be characterized by the X-ray diffraction using the monochromatic Cu K$\alpha$ radiation\cite{Takizawa}.
Then, the strains could be generated by a pair of diamond anvils\cite{BiTeI_exp}, which was used to generate strong pressure even above 200 GPa\cite{Anvil}. Moreover, the real-time pressure strength could be detected by ruby fluorescence method\cite{CQJin1,BiTeI_exp}.
In order to verify the topological property of the material, it is suggested to perform the transport measurements\cite{Hamlin}. Similar to Bi$_2$Se$_3$, the observation of the spin-Hall current\cite{Dora} and the non-equally spaced Landau levels\cite{QKXue1} in IrBi$_3$ will be signatures of the Dirac fermions in surface of the topological insulator\cite{BHZ}.

%In conclusion, we predicted a new kind of material, bismuth based skutterudite, and verified its stability.
In this work, we predict a d-p band inversion topological insulator bismuth-based skutterudite IrBi$_3$, and verify its stability.
Our results indicate that this material is zero gap semi-metal after imposing uniform strain, and it can become topological insulator if an anisotropy is further applied to break the cubic symmetry.
%Moreover, Direct $Z_2$ calculation as well as surface states calculation confirmed its topological character.
Furthermore, near the Fermi level there is a large proportion of d-electronic states which is distinctive with usual topological insulators, for instance in Bi$_2$Se$_3$ and BiTeI the bands involved in the topological band-inversion process are only p-orbitals. Consequently, the electronic interaction in this topological insulator is much stronger than that in other conventional topological insulators.
This provides realistic material for investigating the effect of correlations on the topology, fabricating quantum information devices and spintronics devices with higher stability.

\section*{Methods}

%\noindent \textbf{Details of the DFT calculations}
%\noindent \textbf{Details of the electronic structure calculation}
Our first principle calculations are in the framework of the generalized gradient approximation (GGA) of the density functional theory. The VASP package\cite{VASP1,VASP2}  has been employed and the projector-augmented-wave pseudo-potentials \cite{PAW} are used. Plane waves with a kinetic energy cutoff E$_c$ of 400 eV are used as basis sets and k-point grids in Brillouin zone is chosen as 6$\times$6$\times$6 according to the Monkhorst-Pack scheme. During the ionic relaxation process, the conjugate gradient algorithm is utilised.
In the finite temperature molecular dynamics simulations, a $2\times2\times2$ supercell containing 256 atoms is used and the length of time-step is chosen as 5 fs. The phonon dispersion curve and phonon density of states are obtained using the force-constant method by phonopy code\cite{phonopy}.
 %In the calculation of IrBi$_3$, the crystal structure of the CoSb$_3$ was used while the lattice constant is determined self-consistently as 9.806{\AA}, a bit larger than 9.253{\AA} of IrSb$_3$.
The effect of spin-orbit coupling (SOC) is included in the calculations after the structural relaxations. GGA+U calculations are based on the Dudarev's approach implemented in VASP, with the effective on site Coulomb interaction parameter $U=3.0$ eV and the effective on site exchange interaction parameter $J=0.5$ eV for d-orbitals of Ir atoms\cite{Solovyev}. We also use the modified Becke-Johnson (mBJ) semilocal exchange-correlation potential\cite{mBJ_JCP,mBJ_TranBlaha} to further check the band order and the magnitude of energy gap, and in this process the GGA wave function is used to initialize the mBJ calculation.

%   Nature Communications uses standard Nature referencing style.
%   All authors should be included in reference lists unless there are six or more,
%   in which case only the first author should be given, followed by 'et al.'.
%   Authors should be listed last name first, followed by a comma and initials
%   (followed by full stops) of given names.
%   Article titles should be in Roman text,
%   only the first word of the title should have an initial capital
%   and the title should be written exactly as it appears in the work cited,
%   ending with a full stop.
%   Book titles should be given in italics and all words in the title
%   should have initial capitals. Journal names are italicized and abbreviated
%   (with full stops) according to common usage.
%   Volume numbers and the subsequent comma appear in bold.
%   The full page range should be given, where appropriate.

%\section*{References}

\begin{addendum}

\item [Acknowledgement]
We acknowledge helpful discussions with H. M. Weng and X. X. Wu.
This work is supported by the NKBRSFC (Grants Nos. 2011CB921502, 2012CB821305),
 NSFC (grants Nos. 61227902, 61378017).
The numerical calculations are performed on the Shenteng supercomputer at CNIC-CAS and on the Dawning cluster at IOP-CAS.

\item [Author Contributions]
M. Y. performed the numerical calculations.
All authors analyzed the data and wrote the manuscript.

\item [Competing Interests]
The authors declare that they have no competing financial interests.

\item [Correspondence]
Correspondence and requests for materials should be addressed to Wu-Ming Liu.
\end{addendum}

\clearpage

\newpage
\bigskip
\textbf{Figure 1 Crystal structure and Brillouin zone.}
(a) unit cell of IrBi$_3$, including 8 Ir atoms (green) and 24 Bi atoms (pink). Each Ir atom is surrounded by 6 Bi atoms and each Bi atom has 2 Ir nearest neighbors. (b) The equivalent primitive cell of IrBi$_3$, containing 4 Ir (green) atoms and 12 Bi atoms (pink). (c) The corresponding Brillouin zone and high symmetric points with $\Gamma$ (0,0,0), H (0,1/2,0), N (1/4,1/4,0), P (1/4,1/4,1/4). (d) Free energy as a function of lattice constant.

\bigskip
\textbf{Figure 2 Phonon dispersion and phonon density of states for IrBi$_3$.}
Orange dotted lines in all subfigures denotes the zero frequency. Calculations are performed at zero strain. (a) phonon dispersion curves for IrBi$_3$, in which the inset shows the dispersion near the zero energy. (b) phonon density of states for IrBi$_3$, in which black solid line represents the total phonon density of states, while the green and red shaded areas represent the states coming from Ir and Bi atoms, respectively. Phonon states in the low energy range are mostly composed of states of Bi atoms, indicating that Bi atoms in IrBi$_3$ are much easier to vibrate than the Ir atoms. The phonon dispersion and phonon density of states shows no imaginary frequency, indicating that IrBi$_3$ is stable.

\bigskip
\textbf{Figure 3 Finite temperature molecular dynamics.}
Free energies as functions of time-step at temperature T=30 K (blue curve) and T=300 K (red curve). The slight shift of the free energy curves corresponds to the oscillations of each atom around their equilibrium position. The absence of sharp changes in such curves indicates that no structural phase-transition happens throughout the whole simulation process.
The initial crystal structure (denoted by the orange circle on the free energy curve) is plotted in inset (a). The crystal structure corresponding to the last free energy maximum (denoted by the green circle on the free energy curve) is shown in inset (b) as a comparison. It can be seen that, the latter still shows no significant structural differences as compared with the initial crystal structure.

\bigskip

\textbf{Figure 4 Band structures of IrBi$_3$.}
The black and blue lines in all subfigures represent the GGA band structures and GGA+U band structures respectively. (a) band structure without exerting pressure, the system is in normal metal state with its bands go across the Fermi level several times. (b) band structure under 9\% uniform strain, a zero gap metal state is obtained. (c) further impose a 2\% suppression on the length of c-axis of the primitive cell, a gap appeared at the Fermi level due to the breaking of the cubic symmetry. The inset of (c) is the zoom-in of the band structure close to the Fermi level. (d) Ir-d projected band structure near Fermi level, the radii of red circles are proportional to the weight of Ir-d states, showing a significant band inversion .

\bigskip

\textbf{Figure 5 The atomic- and orbital-resolved density of states.}
The black solid lines in all subfigures represent the total density of states (DOS). (a) atomic-resolved DOS, in which the green curve represents the states of Ir and the red curve represents the states of Bi. It's clear that both type of atom made a significant contribution to the total DOS, different from MoS$_2$ where states near Fermi level are dominated by Mo.  (b) and (c) are orbital-resolved DOS of Ir and Bi atom respectively. Green, blue and red curves represent s-, p- and d-orbitals.

\bigskip
\textbf{Figure 6 Orbital-projected band structures.}
The orange, violet, red, green and blue colors in subfigures represent the d$_{z^2}$, d$_{x^2-y^2}$ and d$_{xy}$, d$_{yz}$ and d$_{xz}$ orbitals respectively. The radii of circles are proportional to the weights of corresponding orbitals. The Fermi level is set to be zero energy. It can be seen that, the t$_{2g}$ orbitals (including the d$_{xy}$, d$_{yz}$ and d$_{xz}$ orbitals) reside far below the Fermi level and are fully occupied. While, the lowest three conduction bands are mainly contributed by the e$_g$ orbitals (including the d$_{z^2}$ and d$_{x^2-y^2}$ orbitals). More specifically, the d$_{x^2-y^2}$ orbital makes an even larger contribution than the the d$_{z^2}$ orbital for the lowest conduction band.
\bigskip
%
%\textbf{Figure 4 Surface states.}
%surface states corresponding to a supercell of N=19 layers by Most Localized Wannier Function(MLWF) approach.   $\Sigma_1$ band crossing the Fermi level 1 times in half BZ is topological surface state.
%%while other ($\Sigma_2$,$\Sigma_3$ and $\Sigma_4$) the topological trivial surface states caused by dangling bonds.

%\iffalse
%\newpage
%\begin{figure}
%\begin{minipage}[t]{0.5\linewidth}
%\centering
%\includegraphics[width=1.7in]{1_a_convensionalcell.eps}
%\end{minipage}%
%\begin{minipage}[t]{0.5\linewidth}
%\centering
%\includegraphics[width=1.7in]{1_b_simplecell.eps}
%\end{minipage}
%\begin{minipage}[t]{0.5\linewidth}
%\centering
%\includegraphics[width=1.7in]{1_c_BZ.eps}
%\end{minipage}%
%\begin{minipage}[t]{0.5\linewidth}
%\centering
%\includegraphics[width=2.0in]{1_d_free_energy.eps}
%\end{minipage}
%\end{figure}
\newpage
\begin{figure}
\begin{center}
\epsfig{file=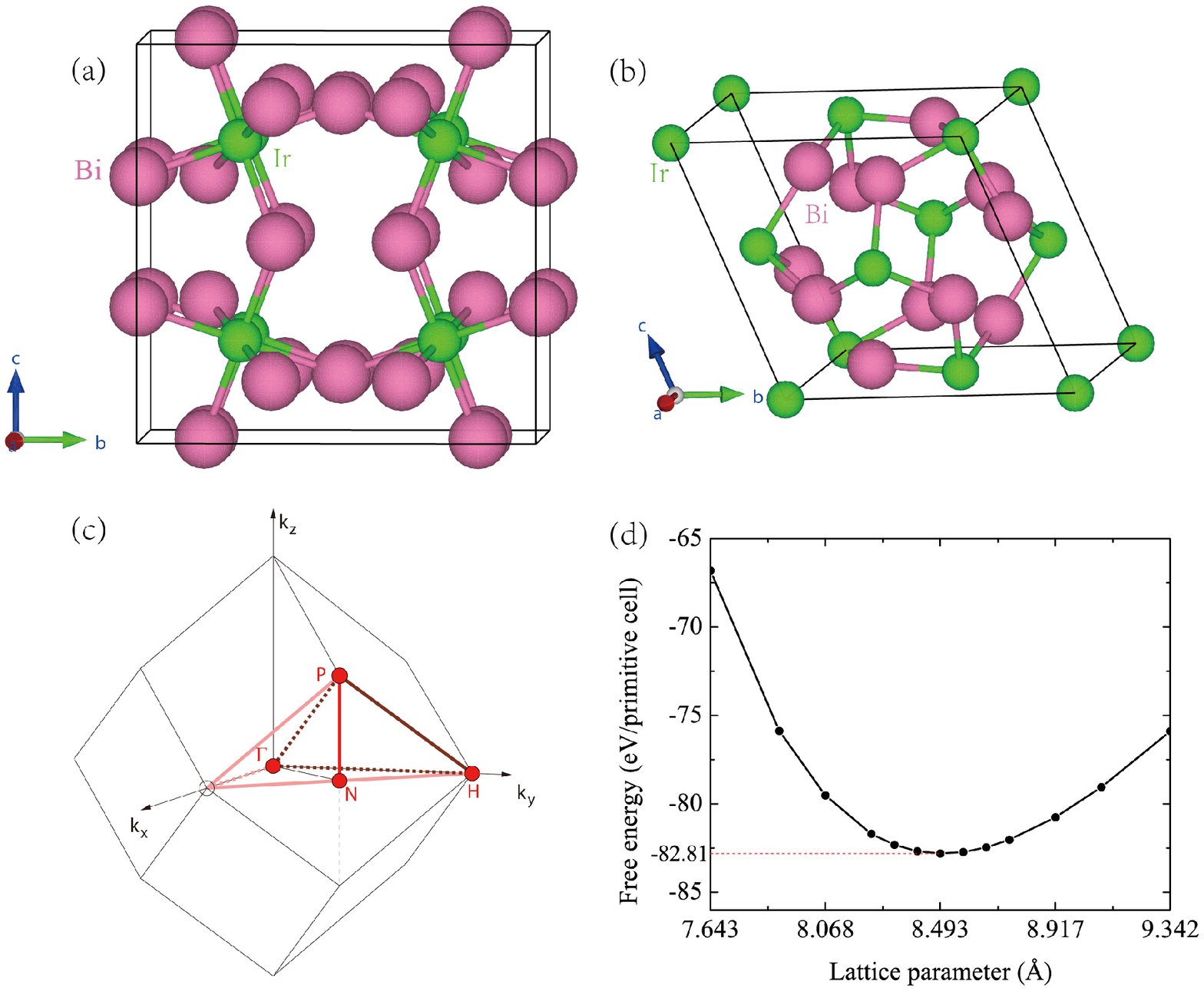,width=15cm}
\end{center}%\caption{}
\label{fig:structure}
\end{figure}

\begin{figure}
\begin{center}
\epsfig{file=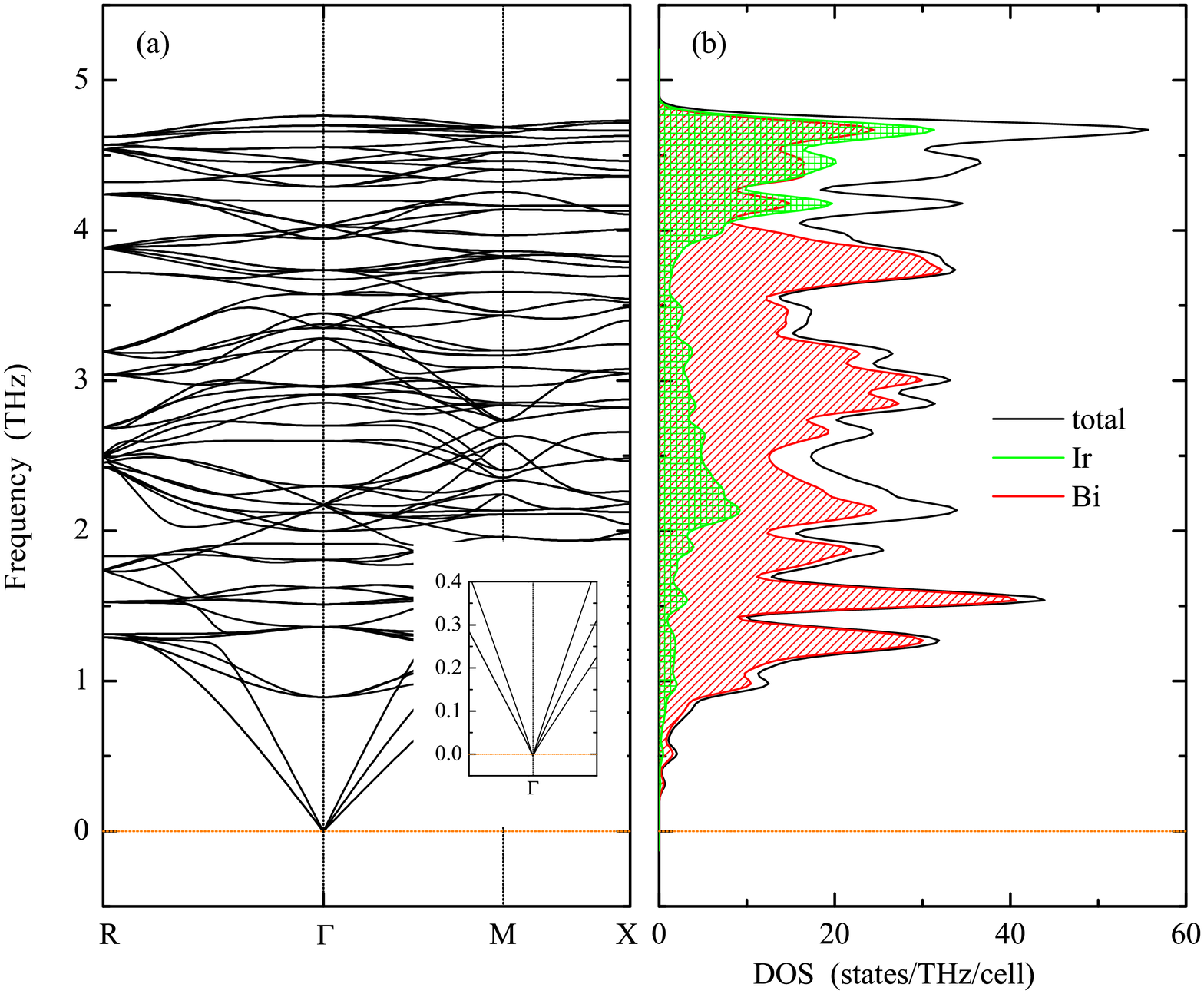,width=15cm}
\end{center}%\caption{}
\label{fig:pho_band_DOS}
\end{figure}

\begin{figure}
\begin{center}
\epsfig{file=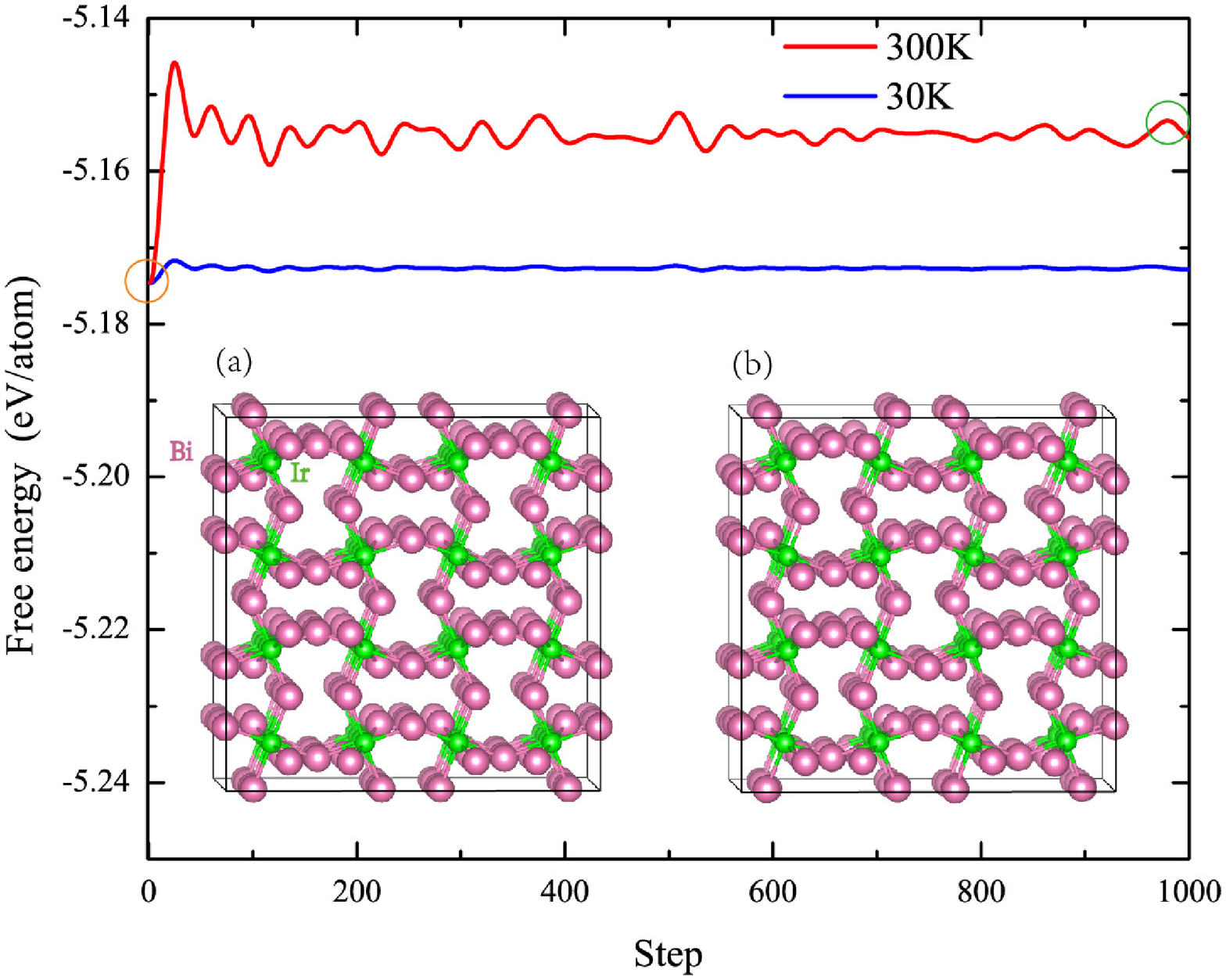,width=13cm}
\end{center}%\caption{}
\label{fig:FTMD}
\end{figure}

\begin{figure}
\begin{center}
\epsfig{file=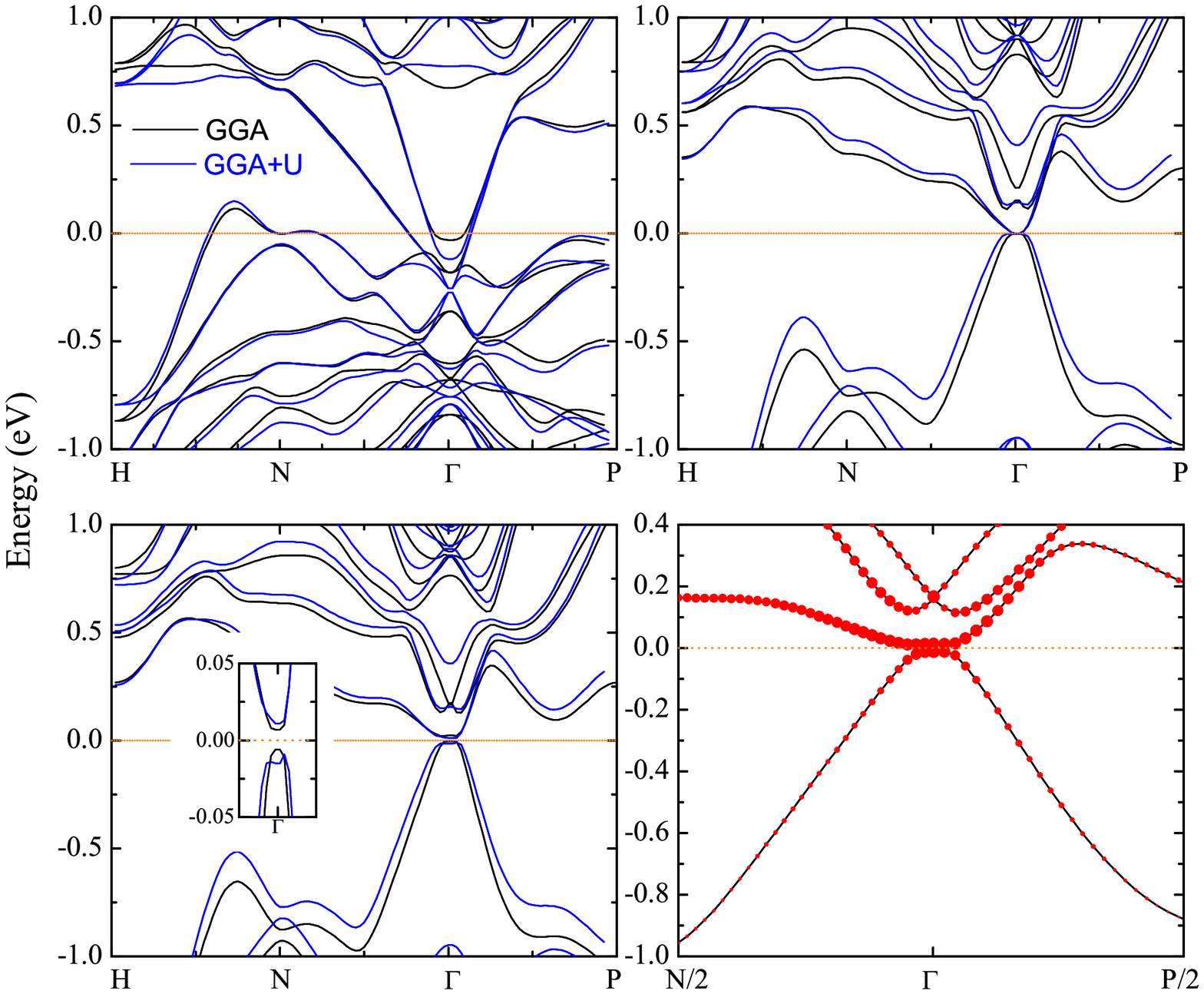,width=13cm}
\end{center}%\caption{}
\label{fig:band}
\end{figure}

\begin{figure}
\begin{center}
\epsfig{file=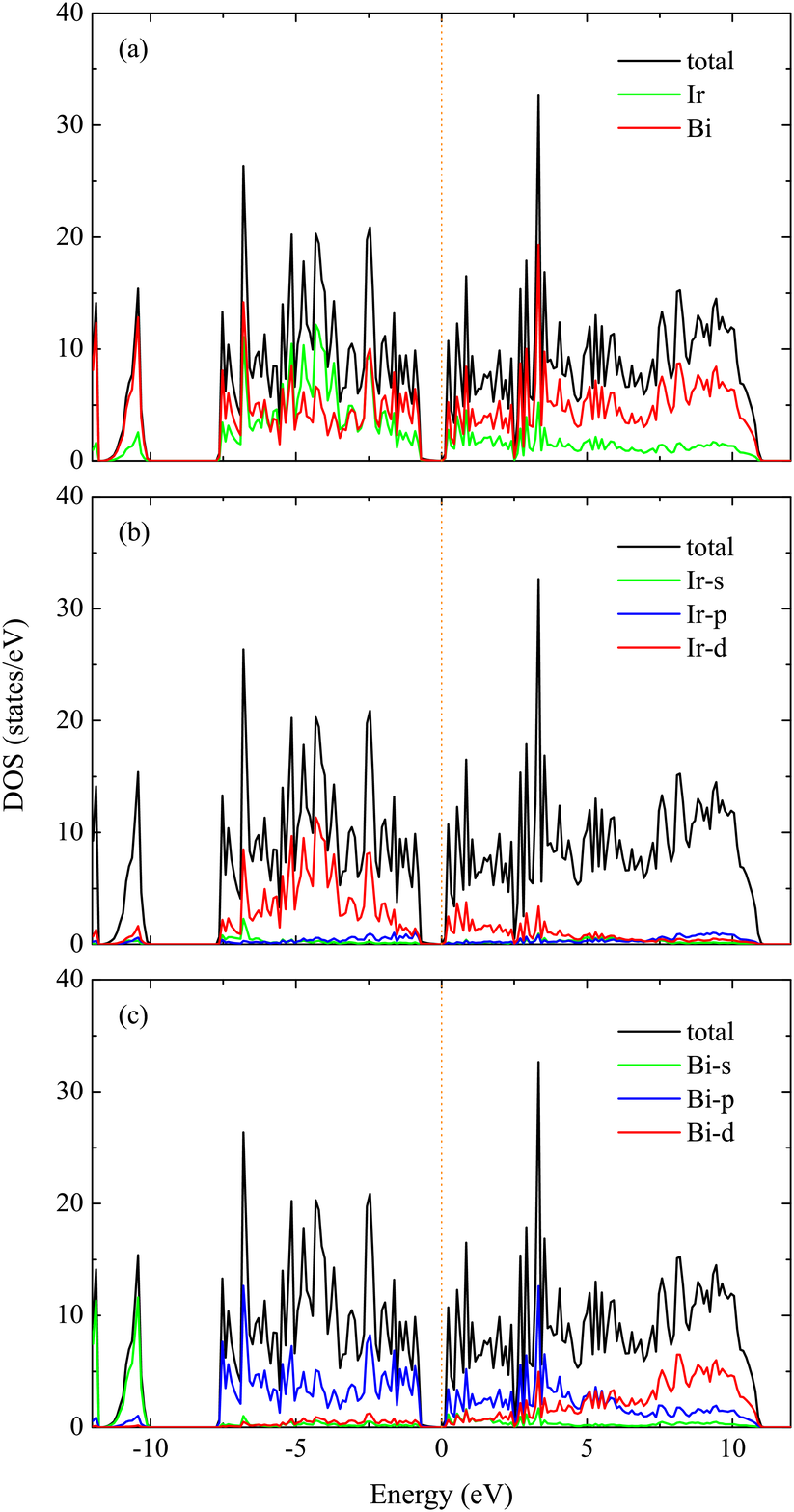,width=12cm}
\end{center}%\caption{}
\label{fig:PDOS}
\end{figure}
\bigskip

\begin{figure}
\begin{center}
\epsfig{file=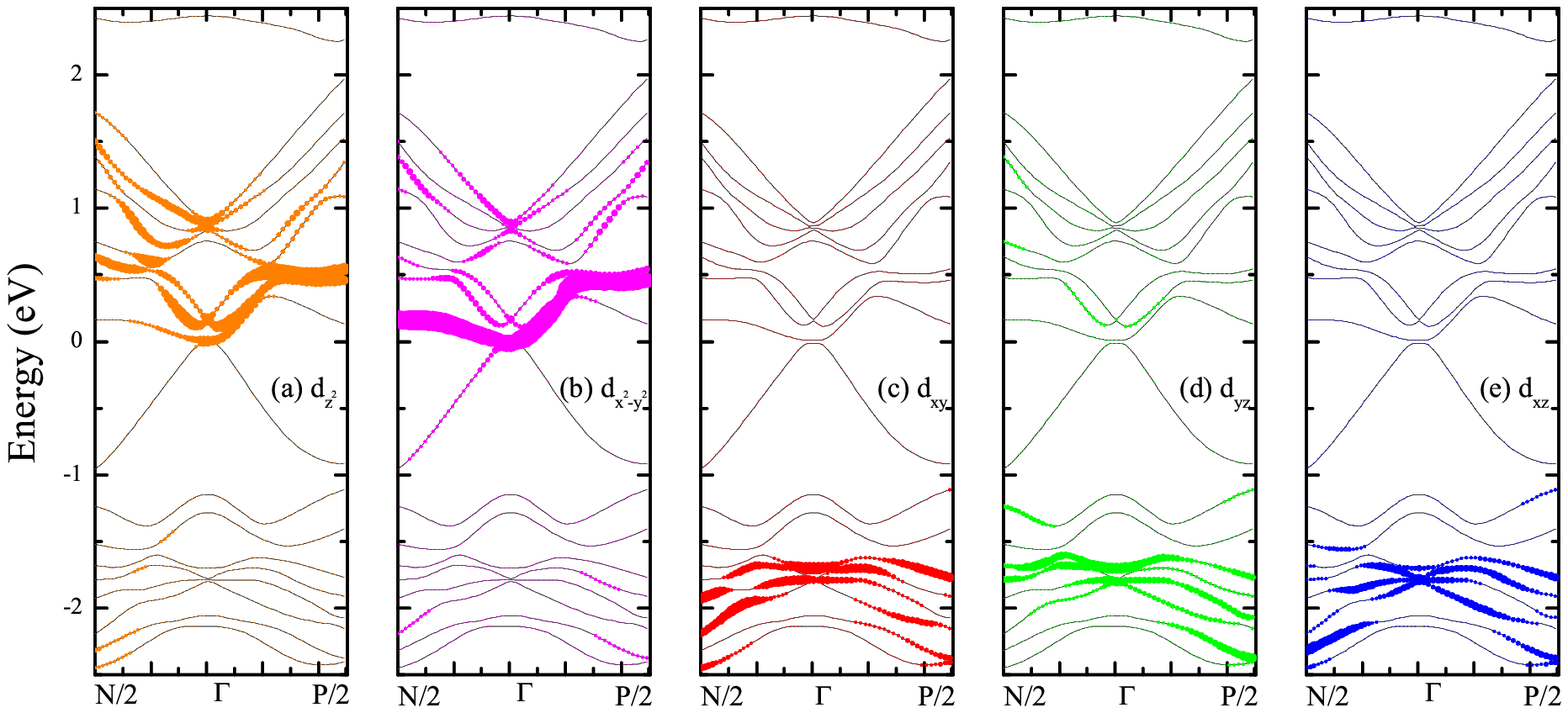,width=16cm}
\end{center}
\label{fig:5d_proj}
\end{figure}
\bigskip

\newpage
\begin{table*}[ht]
\caption{Parities of top-most isolated valence bands at eight time-reversal invariant momenta. Positive parity is denoted by $+$ while negative denoted by $-$. Products of the occupied bands at each time-reversal invariant momentum are listed in the right-most column. As is shown, the product of parities of occupied bands contributes a $-1$ at (0,0,0) while $+1$ at the seven other time-reversal invariant momenta, resulting in $\nu_{0}=1$, $\nu_{1}=\nu_{2}=\nu_{3}=0$.  } % title of Table
\begin{tabular}[t]{c|c|c}
\hline
(0,0,0)&$+++++++++-+--++++++++---+++--++-+--+$&\textbf{$-$}\\
($\pi$,0,0)&$-+-+-+-+-++--+-+-++--+-+-+-++--++-+-$&+\\
(0,$\pi$,0)&$+-+--++-+--++-+-+--++-+-+-+--++--+-+$&+\\
(0,0,$\pi$)&$-+-++--+-++--+-+-++---++-+-++--++-+-$&+\\
($\pi$,$\pi$,0)&$+-+--++-+--++-+-+--++-+-+-+--++--+-+$&+\\
($\pi$,0,$\pi$)&$-+-+-+-+-++--+-+-++--+-+-+-++--++-+-$&+\\
(0,$\pi$,$\pi$)&$+-+--++-+--++-+-+--++-+-+-+--++-+--+$&+\\
($\pi$,$\pi$,$\pi$)&$+++++++-++--+++-++-+-+++-+--++++--+-$&+\\
\hline
\end{tabular}
\label{table:parities} % is used to refer this table in the text
\end{table*}
%
%
%%\clearpage
%\begin{figure}
%\begin{center}
%\epsfig{file=Figure-2.eps,width=14cm}
%\end{center}%\caption{}
%\label{fig:ES}
%\end{figure}
%
%%\clearpage
%\begin{figure}
%\begin{center}
%\epsfig{file=Figure-3.eps,width=15cm}
%\end{center}%\caption{}
%\label{fig:TQPT}
%\end{figure}
%
%%\clearpage
%\begin{figure}
%\begin{center}
%\epsfig{file=Figure-4.eps,width=15cm}
%\end{center}%\caption{}
%\label{fig:FiniteT}
%\end{figure}
%\clearpage
%\bigskip
%%\fi

\newpage
%\title{Supplementary information for ``The d-p band inversion topological insulator in bismuth-based skutterudites''}

\begin{flushleft}
\Large
Supplementary information for ``The d-p band inversion topological insulator in bismuth-based skutterudites''
\normalsize

Ming Yang and Wu-Ming Liu$^{\star}$
\end{flushleft}
\maketitle

\begin{affiliations}
\item
Beijing National Laboratory for Condensed Matter Physics,
Institute of Physics, Chinese Academy of Sciences,
Beijing 100190, China

$^\star$e-mail: wliu@iphy.ac.cn

\end{affiliations}

\subsection{Movies of the evolution of atomic positions in the finite temperature molecular dynamics simulations at temperature 30 K and 300 K.}
In the movies, the Ir atoms and the Bi atoms are denoted by blue balls and green balls, respectively.  A $2\times2\times2$ supercell containing 256 atoms is used in the finite temperature molecular dynamics simulations. It can be seen from the movies that, the atoms shake around their equilibrium positions back and forth while the extent of such motion under 300 K is larger than under 30 K. However, no structural collapse happens throughout the simulations, indicating that IrBi$_3$ is dynamically stable both at low temperature 30 K and at room temperature 300 K.

\subsection{Verification of the band structures with the modified Becke-Johnson potential.}
Here we verify the aforementioned generalized gradient approximations (GGA) band-structures by using the modified Becke-Johnson (mBJ) potential, which is proved to be able to predict an accurate band gap and band order. Firstly, we consider how the electronic density of states changes in mBJ case as compare with that in GGA case. Figure S1 shows the density of states of IrBi$_3$ at its equilibrium status with GGA (blue curve) and mBJ (red curve) potentials. It can be clearly seen that, mBJ potential made a significant modification on the density of states. The conduction bands shift towards higher energy direction. This indicates that, the isotropic strain needed to produce a topological state is reduced, as compared with the results obtained by GGA. Actually, the mBJ result shows IrBi$_3$ with 8\% isotropic strain and 2\% c-axis suppression has already become topological insulator with its band-structure shown in figure S2 (a). The inset of figure S2 (a) shows the zoom-in of the band-structure at the vicinity of Fermi level, from which the band gap is estimated to be 20 meV, corresponding to temperature 232 K which is much higher than the liquid nitrogen temperature.
Figure S2 (b) is the orbital-projected band-structure, in which the radii of red (blue) circles represent the weight of the Ir-d (Bi-p) orbitals. It's clear that the top-most valence band is contributed mostly from the p-orbitals of Bi atoms away from $\Gamma$ point. However, in the vicinity of $\Gamma$ point, this band posses a large weight of d-orbitals of Ir atoms. This shows a obvious band-inversion process between d- and p-orbitals. Such d-p band-inversion character is distinctively different from usual TI materials, for example in HgTe the topological band-inversion happens between s- and p-orbitals and in Bi$_2$Se$_3$ the topological band-inversion happens between p- and p-orbitals.
%states near $E_F$ mainly containing s-p electrons in HgTe and p-electrons in Bi$_2$Se$_3$. These above listed results obtain by mBJ pseudo-potential are consistent with the GGA results.
%As is shown in figure S2 (a), the band-structure of

\newpage
\bigskip
\begin{figure}
\begin{center}
\epsfig{file=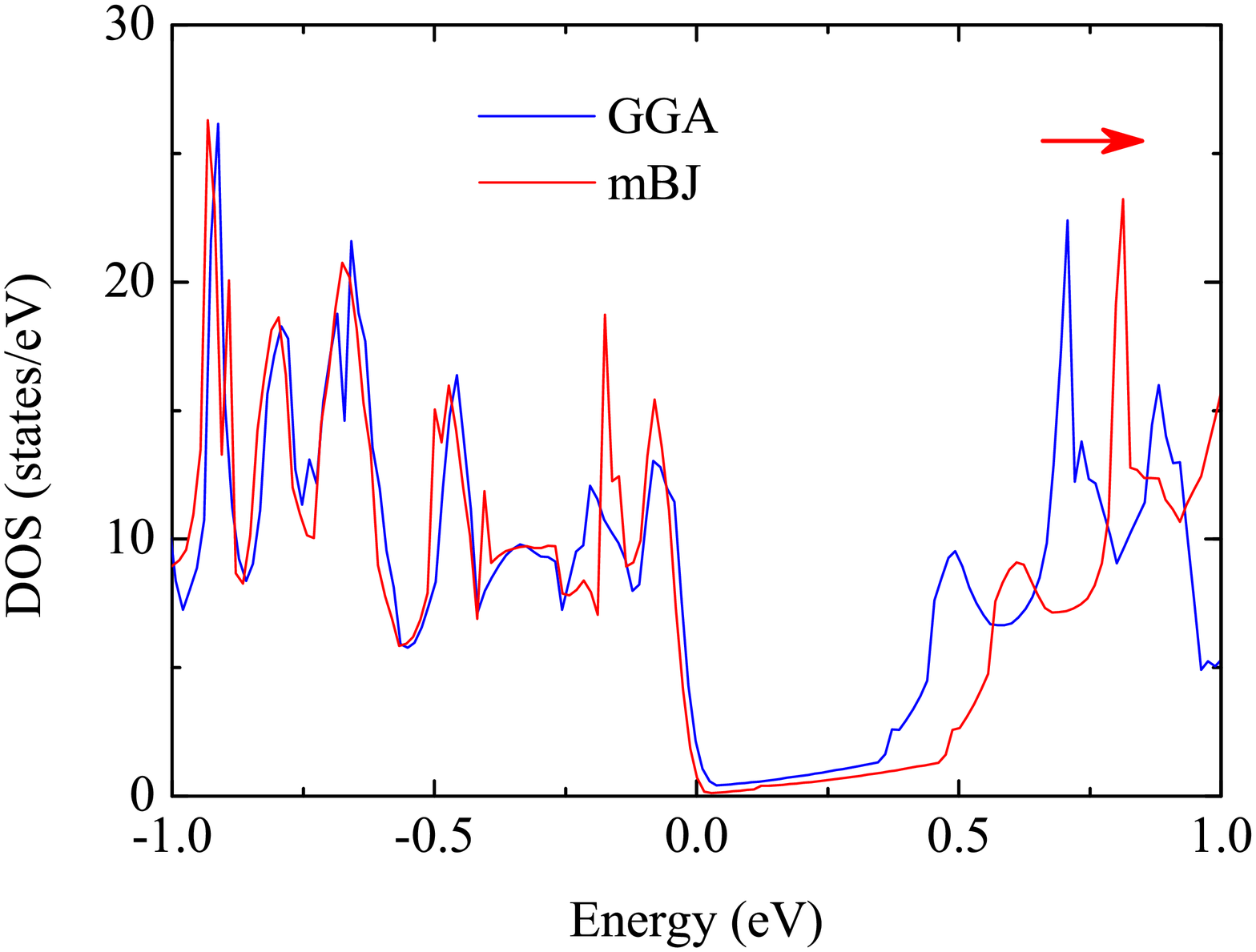,width=15cm}
\end{center}%\caption{}
\label{fig:structure}
\end{figure}
\textbf{Figure S1 The density of states of IrBi$_3$ with GGA pseudo-potential (blue curve) and the modified Becke-Johnson potential (red curve).}
It can be seen that, the modified Becke-Johnson (mBJ) potential made a significant modification on the density of states. The conduction bands shift towards higher energy direction, as is instructed by the red arrow. This indicates that, the isotropic strain needed to produce a topological state is reduced, as compared with the results obtained by GGA.

\newpage
\bigskip
\begin{figure}
\begin{center}
\epsfig{file=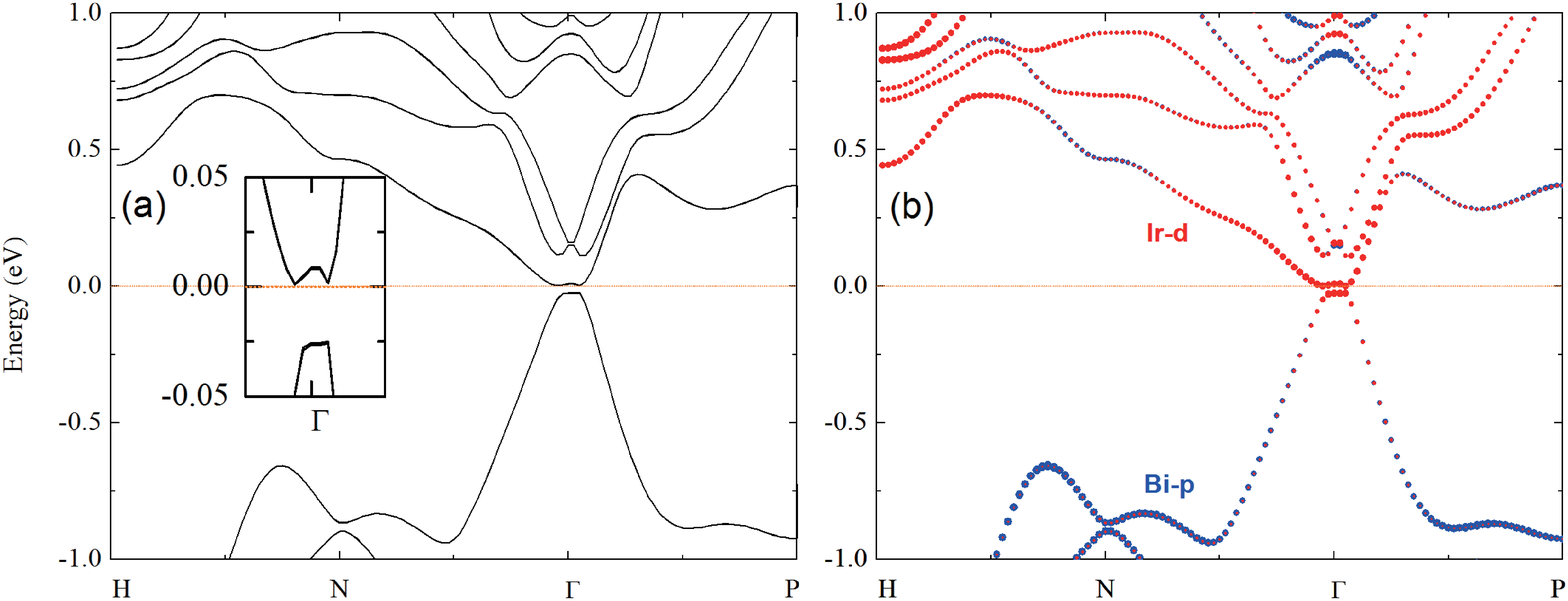,width=15cm}
\end{center}%\caption{}
\label{fig:structure}
\end{figure}
\textbf{Figure S2 band-structure of IrBi$_3$ with 8\% isotropic strain and 2\% c-axis suppression by the modified Becke-Johnson potential.}
Subfigure (a) is the band-structures of IrBi$_3$ with 8\% isotropic strain and 2\% c-axis suppression.
From the inset we can read the gap at the Fermi level is 20 meV, corresponding to temperature 232 K which is much higher than the liquid nitrogen temperature.
Subfigure (b) is the orbital-projected band-structure, in which the radii of red (blue) circles represent the weight of the Ir-d (Bi-p) orbitals. It's clear that the top-most valence band is contributed mostly from the p-orbitals of Bi atoms away from $\Gamma$ point. However, in the vicinity of $\Gamma$ point, this band posses a large weight of d-orbitals of Ir atoms. This shows a obvious band-inversion process between d- and p-orbitals. Such d-p band-inversion character is distinctively different from usual TI materials, for example in HgTe the topological band-inversion happens between s- and p-orbitals and in Bi$_2$Se$_3$ the topological band-inversion happens between p- and p-orbitals.


\begin{thebibliography}{99}

\bibitem{BHZ} Bernevig, B. A., Hughes, T. A. \& Zhang, S. C.
Quantum spin Hall effect and topological phase transition in HgTe quantum wells.
\textit{Science} \textbf{314}, 1757 (2006).

\bibitem{Heusler_SCZhang} Chadov, S., Qi, X. L., Kubler, J., Fecher, G. H., Felser, C., Zhang S. C.
Tunable multifunctional topological insulators in ternary Heusler compounds.
\textit{Nat. Mat.} \textbf{9}, 541 (2010).

\bibitem{magnetoelectric} Qi, X. L., Hughes, T. L. \& Zhang, S. C.
Topological field theory of time-reversal invariant insulators.
\textit{Phys. Rev. B} \textbf{78}, 195424 (2008).

\bibitem{Fang_Bi2Se3} Zhang, H. J., Liu, C. X., Qi, X. L., Dai, X., Fang Z. \& Zhang, S. C.
Topological insulators in Bi2Se3, Bi2Te3 and Sb2Te3 with a single Dirac cone on the surface.
\textit{Nat. Phys.} \textbf{5}, 438 (2009).

\bibitem{Fu} Fu, L. \& Kane, C. L.
Topological insulators with inversion symmetry.
\textit{Phys. Rev. B} \textbf{76}, 045302 (2007).

\bibitem{YXia} Xia, Y. \textit{et al}.
Observation of a large-gap topological-insulator class with a single Dirac cone on the surface.
\textit{Nat. Phys.} \textbf{5}, 398 (2009).

\bibitem{Yan1} Yan, B. H., Muchler, L., Qi, X. L., Zhang, S. C. \& Felser, C.
Topological insulators in filled skutterudites,
\textit{Phys. Rev. B} \textbf{85}, 165125 (2012).

\bibitem{silicene_Yao} Liu, C. C., Feng, W. X. \& Yao, Y. G.
Quantum spin Hall effect in silicene and two-dimensional germanium.
\textit{Phys. Rev. Lett.} \textbf{107}, 076802 (2011).

\bibitem{SunFD} Sun, F. D., Yu, X. L., Ye, J. W., Fan, H. \& Liu, W. M.
Topological quantum phase transition in synthetic non-abelian gauge potential: gauge invariance and experimental detections.
\textit{Scientific Reports} \textbf{3}, 2119 (2013).

\bibitem{QAHE} Yu, R., Zhang, W., Zhang, H. J., Zhang, S. C., Dai, X. \& Fang, Z.
Quantized anomalous Hall effect in magnetic topological insulators.
\textit{Science} \textbf{329}, 61 (2010).

\bibitem{Ding} Ding, J., Qiao, Z. H., Feng, W. X., Yao, Y. G. \& Niu, Q.
Engineering quantum anomalous/valley Hall states in graphene via metal-atom adsorption: An ab-initio study.
\textit{Phys. Rev. B} \textbf{84}, 195444 (2011).

\bibitem{ZHQiao} Qiao, Z. H., Tse, W., Jiang, H., Yao, Y. G. \& Niu, Q.
Two-Dimensional topological insulator state and topological phase transition in bilayer graphene.
\textit{Phys. Rev. Lett.} \textbf{107}, 256801 (2011).
%
%\bibitem{WuCJ} Liu, X. J., Liu, X., Wu, C. J., Sinova, J.
%Quantum anomalous Hall effect with cold atoms trapped in a square lattice.
%\textit{Phys. Rev. A} \textbf{81}, 033622 (2010).
%
%\bibitem{TCI_Tanaka} Tanaka, Y. \textit{et al}.
%Experimental realization of a topological crystalline insulator in SnTe.
%\textit{Nat. Phys.} \textbf{8}, 800 (2012).
%
%\bibitem{TCI_Dziawa} Dziawa, P. \textit{et al}.
%Topological crystalline insulator states in Pb1-xSnxSe.
%\textit{Nat. Mat.} \textbf{11}, 1023 (2012).
%
%\bibitem{TCI_Hsieh2} Hsieh, T. H. \textit{et al}.
%Topological crystalline insulators in the SnTe material class.
%\textit{Nat. Commun.} \textbf{3}, 679 (2012).
%
%\bibitem{TCI_Hsieh1} Hsieh, T. H. \textit{et al}.
%Observation of a topological crystalline insulator phase and topological phase transition in Pb1-xSnxTe.
%\textit{Nat. Commun.} \textbf{3}, 1192 (2012).
%
%\bibitem{GaS} Zhu, Z., Cheng, Y. \& Schwingenschlogl, U.
%Topological phase transition in layered gas and gase.
%\textit{Phys. Rev. Lett.} \textbf{108}, 266805 (2012).


\bibitem{ZhangXL} Zhang, X. L., Liu, L. F. \& Liu, W. M.
Quantum anomalous Hall effect and tunable topological states in 3d transition metals doped silicene.
\textit{Scientific Reports} \textbf{3}, 2908 (2013).

\bibitem{JianminZhang} Zhang, J. M., Zhu, W. G., Zhang, Y., Xiao, D. \& Yao, Y. G.
Tailoring magnetic doping in the topological insulator Bi2Se3.
\textit{Phys. Rev. Lett.} \textbf{109}, 266405 (2012).

\bibitem{Majorana_Fu} Fu, L., \& Kane C. L.
Superconducting proximity effect and Majorana fermions at the surface of a topological insulator.
\textit{Phys. Rev. Lett.} \textbf{100}, 096407 (2008).

\bibitem{Majorana_Tiwari} Tiwari, R. P., Z¨¹licke, U. \& Bruder U.
Majorana fermions from Landau quantization in a superconductor and topological-insulator hybrid structure.
\textit{Phys. Rev. Lett.} \textbf{100}, 186805 (2013).

\bibitem{weizhang} Zhang, W., Yu, R., Zhang, H. J., Dai, X. \& Fang, Z.
First-principles studies of the three-dimensional strong topological insulators Bi2Te3, Bi2Se3 and Sb2Te3.
\textit{New Journal of Physics} \textbf{12}, 065013 (2010).

\bibitem{Bahramy} Bahramy, M. S., Yang, B. J., Arita, R. \& Nagaosa, N.
Emergent quantum confinement at topological insulator surfaces.
\textit{Nat. Commun.} \textbf{3}, 679 (2012).

\bibitem{GuoGuangYu} Wang, C. R.  \textit{et al}.
Magnetotransport in copper-doped noncentrosymmetric BiTeI.
\textit{Phys. Rev. B} \textbf{88}, 081104(R) (2013).

\bibitem{Yao_Chalcopyrite} Feng, W. X., Xiao, D., Ding, J. \& Yao, Y. G.
Three-dimensional topological insulators in I-III-VI2 and II-IV-V2 chalcopyrite semiconductors,
\textit{Phys. Rev. Lett.} \textbf{106}, 016402 (2011).

\bibitem{xiangtan} Liu, W. L. \textit{et al}.
Anisotropic interactions and strain-induced topological phase transition in Sb2Se3 and Bi2Se3.
\textit{Phys. Rev. B} \textbf{84}, 245105 (2011).


%\bibitem{LiaoRY} Liao, R. Y., Yu, Y. X. \& Liu, W. M.
%Tuning the tricritical point with spin-orbit coupling in polarized fermionic condensates.
%\textit{Phys. Rev. Lett.} \textbf{108}, 196802 (2012).

\bibitem{Heusler_Yao} Xiao, D. \textit{et al}.
Half-Heusler compounds as a new class of three-dimensional topological insulators.
\textit{Phys. Rev. Lett.} \textbf{105}, 096404 (2010).

\bibitem{XieXC} Yu, S. L., Xie, X. C. \& Li, J. X.
Mott physics and topological phase transition in correlated dirac fermions.
\textit{Phys. Rev. Lett.} \textbf{107}, 010401 (2011).

\bibitem{Castro} Castro, E. V. \textit{et al}.
Topological fermi liquids from coulomb interactions in the doped honeycomb lattice.
\textit{Phys. Rev. Lett.} \textbf{107}, 106402 (2011).

%\bibitem{Kim} Kim, M. S., Kim, C. H., Kim, H. S. \& Ihm, J.
%Topological quantum phase transitions driven by external electric fields in Sb2Te3 thin films.
%\textit{PNAS} \textbf{109}, 671 (2012).
%
%\bibitem{KimD} Kim, D. \textit{et al}.
%Surface conduction of topological Dirac electrons in bulk insulating Bi2Se3.
%\textit{Nat. Phys.} \textbf{8}, 459 (2012).
%
%\bibitem{texture_XSY} Xu, S. Y. \textit{et al}.
%Topological phase transition and texture inversion in a tunable topological insulator.
%\textit{Science} \textbf{332}, 560 (2011).
%
%\bibitem{texture_Henk} Henk, J. \textit{et al}.
%Complex spin texture in the pure and Mn-doped topological insulator Bi2Te3.
%\textit{Phys. Rev. Lett.} \textbf{108}, 206801  (2012).


\bibitem{Niwa} Niwa, K. \textit{et al}.
Compression behaviors of binary skutterudite CoP3 in noble gases up to 40 GPa at room temperature.
\textit{Inorg. Chem.} \textbf{50}, 3281 (2011).

\bibitem{RuSb3} Smalley, A., Jespersen, M. L. \& Johnson, D. C.
Synthesis and structural evolution of RuSb3, a new metastable skutterudite compound.
\textit{Inorg. Chem.} \textbf{43}, 2486 (2004).

\bibitem{Caillat} Caillat, T., Fleurial, J. P. \& Borshchevsky, A.
Bridgman-solution crystal growth and characterization of the skutterudite compounds CoSb3 and RhSb3.
\textit{J. Crystal Growth} \textbf{166}, 722 (1996).

\bibitem{annealing} Akasaka, M. \textit{et al}.
Effects of post-annealing on thermoelectric properties of p-type CoSb3 grown by the vertical Bridgman method.
\textit{J. Alloys and Compounds} \textbf{386}, 228 (2005).

\bibitem{Takizawa} Takizawa, H., Miura, K., Ito, M., Suzuki, B., Endo, T.
Atom insertion into the CoSb skutterudite host lattice under high pressure.
\textit{J. Alloys and Compounds} \textbf{282}, 79 (1999).

\bibitem{BiTeI_exp} Xi, X. X. \textit{et al}.
Signatures of a pressure-induced topological quantum phase transition in BiTeI.
\textit{Phys. Rev. Lett.} \textbf{111}, 155701 (2013).

\bibitem{Anvil} Nakamoto, Y. \textit{et al}.
Generation of Multi-megabar pressure using nano-polycrystalline diamond anvils.
\textit{Jpn. J. App. Phys.} \textbf{46}, 640 (2007).

\bibitem{CQJin1} Zhu, J. L. \textit{et al}.
Superconductivity in topological insulator Sb2Te3 induced by pressure.
\textit{Scientific Reports} \textbf{3}, 2016 (2013).

\bibitem{Hamlin} Hamlin, J. J. \textit{et al}.
High pressure transport properties of the topological insulator Bi2Se3.
\textit{J. Phys.: Condens. Matter} \textbf{24}, 035602 (2012).

\bibitem{Dora} Dora, B. \& Moessner, R.
Dynamics of the spin Hall effect in topological insulators and graphene.
\textit{Phys. Rev. B} \textbf{83}, 073403 (2011).

\bibitem{QKXue1} Cheng, P. \textit{et al}.
Landau quantization of topological surface states in Bi2Se3.
\textit{Phys. Rev. Lett.} \textbf{105}, 076801 (2010).

\bibitem{CoSb3} Sales, B. C.,  Mandrus, D. \& Williams, R. K.
Filled skutterudite antimonides: a new class of thermoelectric materials.
\textit{Science} \textbf{272}, 1325 (1996).

\bibitem{PrOs4Sb12_Seyfarth} Seyfarth, G. \textit{et al}.
Multiband superconductivity in the heavy fermion compound PrOs4Sb12.
\textit{Phys. Rev. Lett.} \textbf{95}, 107004 (2005).

\bibitem{PrOs4Sb12_Matsumoto} Matsumoto, M. \textit{et al}.
Exciton mediated superconductivity in PrOs4Sb12.
\textit{J. Phys. Soc. J.} \textbf{73}, 1135 (2004).

\bibitem{Pickett1} Smith, J. C. , Banerjee, S., Pardo, V. \& Pickett, W. E.
Dirac point degenerate with massive bands at a topological quantum critical point.
\textit{Phys. Rev. Lett.} \textbf{106}, 056401 (2011).
%\bibitem{PrOs4Sb12_Sugawara} Sugawara, H. \textit{et al}.
%Fermi surface of the heavy-fermion superconductor PrOs4Sb12.
%\textit{Phys. Rev. B} \textbf{66}, 220504(R) (2002).

\bibitem{NiSb3} Williams, J. M. \& Johnson, D. C.
Synthesis of the new metastable skutterudite compound NiSb3 from modulated elemental reactants.
\textit{Inorg. Chem.} \textbf{41}, 4127 (2002).

\bibitem{VASP1} Kresse, G.  \& Furthmuller, J.
Efficient iterative schemes for ab initio total-energy calculations using a plane-wave basis set.
\textit{Phys. Rev. B} \textbf{54}, 11169 (1996).

\bibitem{VASP2} Kresse, G.  \& Furthmuller, J.
Efficiency of ab-initio total energy calculations for metals and semiconductors using a plane-wave basis set.
\textit{Comput. Mater. Sci.} \textbf{6},  15 (1996).

\bibitem{PAW} Blochl, P. E.
Projector augmented-wave method.
\textit{Phys. Rev. B}, \textbf{50}, 17953 (1994).

\bibitem{IrSb3} Kjekshus, A. \textit{et al}.
Compounds with the skutterudite type crystal structure. III. structural data for arsenides and antimonides.
\textit{Acta Chemica Scandinavica} \textbf{28}, 99 (1974).


\bibitem{BindingEnergy} There are $n_{Ir}=4$ Ir atoms and $n_{Bi}=12$ Bi atoms in an IrBi$_3$ primitive cell. At GGA level, $E_{Ir}=-8.69$ eV for crystalline Ir with space group $FM\bar{3}M$ and $E_{Bi}=-3.70$ eV for crystalline Bi with space group $IM\bar{3}M$. From Fig. 1(d) we read $E_{IrBi_3}=-82.81$ eV. Substituting the above values in Eq.(1), we arrived at the binding energy $E_{b}=-3.65$ eV.



%\bibitem{FTMD} FTMD simulation was carried out using VASP pakage at 300K. Time step was choosen as 5fs. 1000steps were performed and 32 atoms(including 8 Ir atoms and 24 Bi atoms) were involved in the FTMD simulation. Throughout the simulation, we find each atom oscillate around its equilibrium position rather than run away, and the lattice structure is never destroyed.

%
%\bibitem{wannier1}  Marzari, N. \& Vanderbilt, D.
%Maximally localized generalized Wannier functions for composite energy bands.
%\textit{Phys. Rev. B} \textbf{56}, 12847 (1997).
%
%\bibitem{wannier2} Souza, I., Marzari, N. \& Vanderbilt, D.
%Maximally localized Wannier functions for entangled energy bands.
%\textit{Phys. Rev. B} \textbf{65}, 035109 (2001).

\bibitem{mBJ_JCP} Becke, A., \&  Johnson E.
A simple effective potential for exchange.
\textit{J. Chem. Phys.} \textbf{124}, 221101 (2006).

\bibitem{mBJ_TranBlaha} Tran, F. \& Blaha, P.
Accurate band gaps of semiconductors and insulators with a semilocal exchange-correlation potential.
\textit{Phys. Rev. Lett.} \textbf{102}, 226401 (2009).

\bibitem{mBJ_WXFeng} Feng, W. X., Xiao, D., Zhang, Y., \& Yao, Y. G.
Half-Heusler topological insulators: A first-principles study with the Tran-Blaha modified Becke-Johnson density functional,
\textit{Phys. Rev. B} \textbf{82}, 235121 (2010).

\bibitem{wanxiang_feng_MoS2} Xiao, D., Liu, G. B., Feng, W. X., Xu, X. D. \& Yao, W.
Coupled spin and valley physics in monolayers of MoS2 and other group-VI dichalcogenides.
\textit{Phys. Rev. Lett.} \textbf{108}, 196802 (2012).
%
%\bibitem{WangZJ} Wang, Z. J., Weng, H. M., Wu, Q. S., Dai, X. \& Fang, Z.
%Three-dimensional Dirac semimetal and quantum transport in Cd3As2.
%\textit{Phys. Rev. B} \textbf{88}, 125427 (2013).

\bibitem{phonopy} Togo, A. \textit{et al}.
First-principles calculations of the ferroelastic transition between rutile-type and CaCl2-type SiO2 at high pressures.
\textit{Phys. Rev. B} \textbf{78}, 134106 (2008).

\bibitem{Solovyev} Solovyev, I. V. \& Dederichs, P. H.
Corrected atomic limit in the local-density approximation and the electronic structure of d impurities in Rb.
\textit{Phys. Rev. B} \textbf{50}, 16861 (1994).


\end{thebibliography}
\end{document}